\documentclass{revtex4-1}
\usepackage{mathtools}
\usepackage{graphics}
\usepackage{amsmath}
\usepackage{hyperref}
\usepackage[hypcap]{caption}

\begin{document}
\title{Femtosecond-scale Synchronisation of Ultra-Intense Focused Laser Beams}
\date{\today}
\author{D. J. Corvan$^{1}$}
\email{dcorvan01@qub.ac.uk}
\author{W. Schumaker$^{2}$}
\author{J. Cole$^{3}$}
\author{H. Ahmed$^{1}$}
\author{K. Krushelnick$^{2}$}
\author{S. P. D. Mangles$^{3}$}
\author{Z. Najmudin$^{3}$}
\author{D. Symes$^{5}$}
\author{A. G. R. Thomas$^{2}$}
\author{M. Yeung$^{4}$}
\author{M. Zepf$^{1,4}$}
\author{Z. Zhao$^{2}$}
\author{G. Sarri$^{1}$}
\affiliation {$^{1}$School of Mathematics and Physics, Queen’s University Belfast,
BT7 1NN, Belfast, UK} 
\affiliation {$^{2}$Centre for Ultrafast Optical Science, University of Michigan, 
Ann Arbor, Michigan 48109-2099, USA} 
\affiliation {$^{3}$The John Adams Institute for Accelerator Science, Imperial College of Science, Technology and Medicine,
London, SW7 2AZ, UK} 
\affiliation {$^{4}$Helmoltz Institut Jena, Fr\"{o}belstieg 3,07743 Jena, Germany}
\affiliation {$^{5}$Central Laser Facility, Rutherford Appleton Laboratory , Didcot, Oxfordshire, OX11 0QX, UK}

\begin{abstract}
Synchronising ultra-short ($\sim$fs) and focussed laser pulses is a particularly difficult task, as this timescale lies orders of magnitude below the typical range of fast electronic devices. Here we present an optical technique that allows for femtosecond-scale synchronisation of the focal planes of two focussed laser pulses. This technique is virtually applicable to any focussing geometry and relative intensity of the two lasers. Experimental implementation of this technique provides excellent quantitative agreement with theoretical expectations. The proposed technique will prove highly beneficial for the next generation of multiple, petawatt class laser systems.

\end{abstract}

\maketitle

\section{Introduction}\label{sec:Introinf}
Ultra-intense laser matter interactions generate extreme states of matter whose study is of paramount importance for the understanding of high energy density physics \cite{WDM}. Multiple laser beam interactions are progressively attracting the attention of the physics community as they provide a unique scenario for the experimental study of highly non-linear particle-photon interactions \cite{Piazza}, laboratory astrophysics \cite{Labastro}, 
particle acceleration \cite{2beamacc,twobeamLWFA}, particle generation \cite{bellpairs} and the production of the next generation of x-ray sources \cite{ChrisNLTS,MackNLCS}. 

For a meaningful experimental implementation, it is indeed necessary that the high intensity foci of the lasers be spatially overlapped and temporally synchronised with a micron and femtosecond scale precision respectively. While spatial overlap is relatively easy to achieve, femtosecond scale synchronisation is a much harder task, since it lies orders of magnitude below the typical resolution of electronic devices (which respond on the ns to $100$'s of ps scale). 
 
Previously, it was shown that by using the relatively broad frequency envelope of a femtosecond laser pulse, detailed information about the phase front and relative time delay between two collimated laser beams could be revealed by employing an interferometric technique provided prior knowledge of a reference beam parameters was previously obtained \cite{specinforig,specinfpaper}. Exploiting similar physical principles, we report here on a compact and versatile experimental technique that allows for the fine synchronisation of the focal planes of two focussed laser pulses. The proposed technique can be implemented in virtually any focusing geometry and relative intensity of the two laser beams. The technique has been tested using the two beam laser system Astra-Gemini at the Rutherford Appleton Laboratory (RAL) in Oxford UK, giving quantitative agreement with theoretical expectations.

This technique will prove highly beneficial for the next class of multi-petawatt laser systems such as the $20$ PW Vlucan at RAL \cite{VulcanPW} and the Extreme Light Infrastructure (ELI) \cite{ELI}, in which experiments with multiple ultra-high intensity lasers will be routinely performed. 

The structure of this paper is as follows. In Section \ref{sec:Theoryspecinf} the theory underlying this technique will be discussed and a prediction of the outcome is provided based on data specific to RAL. Then in Section \ref{sec:Expsetinf}, the implementation of the technique is outlined for a recent experimental campaign in Astra-Gemini. Section \ref{sec:resultsinf} presents the experimental findings and a comparison with the predictions is made. Finally, Section \ref{sec:conclusioninf} gives a conclusive overview of the technique.

\section{Theory}\label{sec:Theoryspecinf}

Let us assume two pulses that have an optical path difference $d$ between them (where $d=ND$, $D$ being the separation of the sources and $N$ the refractive index of the material separating them). An interference pattern will be observed with a number of fringes $n$, which is given by $n=2d/\lambda$. The spatial fringes will be observed parallel to the axis of polarisation. Looking at the phase difference $\vartheta$ at each point on this axis, one extracts the equation $\vartheta=2\pi ny$. Here $y$ is the vertical distance on the axis normalised to the total length of that axis.

If the pulses are also incident on a diffraction grating in such a way that the $1^{st}$ order reflection disperses the frequencies along the vector perpendicular to the axis of polarisation, temporal fringes will manifest perpendicular to the spatial fringes. Previously, it was shown that spectral interference of two collinear pulses delayed in time by $\tau$ with respect to each other, they would be observed with a phase difference given by $\varphi_1 - \varphi_2 -\omega \tau $ (where $\varphi$ represents the phase of beams $1$ and $2$ respectively) \cite{specinfpaper}. From the $\omega \tau $ term, it can be seen that for the same change in time delay, different frequencies decouple from one another by different amounts and because a diffraction grating simultaneously resolves frequencies and stretches their time durations, the pulses can interfere even though they are multiple time periods apart. 

At the peak of the laser pulse envelope, the wavelength is given by $\lambda_0$, which has a corresponding time period of $T_0$. Longer wavelengths having a greater time period of $T_+$ and the shorter wavelengths having a smaller time period of $T_-$. The relative difference between these values with respect to the central time period $T_{\pm}/T_0$, will provide information about how these components of the pulses interfere with one another whenever those pulses are subject to a delay. This is achieved by converting the time differences to their corresponding phase shifts on the frequency axis. This shift is revealed by the quantity $\varphi=2\tau T_{\pm}/T_0\times 2\pi$ where $\tau$ is the time periods of delay between the two beams normalised to $T_0$. 

The total phase difference between the two beams at any point on the plane which encompasses the two vectors described above is given by the Matrix $C=\varphi \oplus \vartheta$ i.e., the sum of the phase shifts in spatial axis, and the phase shift in temporal axis at each position on the plane. This is shown in equation \ref{eq:specinfsimp}.
\begin{align}\label{eq:specinfsimp}
I(x,y)\propto\bigg|E_1^2+E_1^2+2E_1E_2\cos\Big[C\Big]\bigg|
\end{align}
Where $E$ denotes the maximum electric field strength of the first laser pulse (subscript $1$) and second laser pulse (subscript $2$). From equation \ref{eq:specinfsimp}, it can be seen that if the electric field of each beam is comparable, then one need only consider the terms within the cosine bracket to describe the changes observed. Therefore, the intensity can be simplified to
\begin{align}\label{eq:specinfsupersimp}
I=\propto\big(1+\cos[C]\big)
\end{align}

Using data from the Astra-Gemini laser system at RAL \cite{Geminispec} where two beams with a duration of $45\pm3$ fs, at a repetition rate of  $10$ Hz, expectations of the behaviour can be determined. The central wavelength of the pulse is $800$ $\sim30$ nm FWHM; hence, the central time period is given as $T_0=2.667\times10^{-15}$ s with $T_-=2.583\times10^{-15}$ s and $T_+=2.750\times10^{-15}$ s. The pellicle used was of thickness $D\sim1.2$ $\mu$m and refractive index $N\sim1.4$ $\mu$m. This corresponds to $d\sim1.68$ and hence, to $\sim4$ spatial fringes. If the axis parallel to the polarisation is assigned to the x axis and the axis perpendicular to the polarisation to the y axis, a prediction of the interference pattern that will be observed can be generated. The expectations are fitted to a window where $x=1.6$ mm and $y=4.5$ mm in order to match the image produced on a CCD camera.

\begin{figure}[h!]
\includegraphics[width=1.0\textwidth]{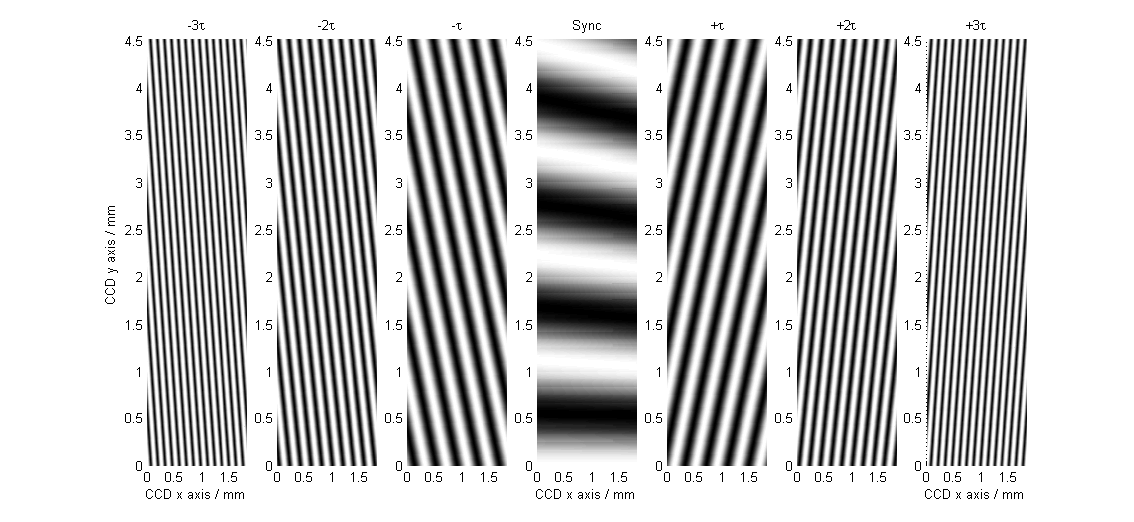} 
  \caption{The relative intensity plotted as a function of the x,y positions with values plotted in term of the phase difference associated with each position on the x-y axis, $\tau$ is the time normalised motion of the delay stage. Moving from left to right the delays are $-3\tau$, $-2\tau$, $-1\tau$, $0$, $+1\tau$, $+2\tau$ and $+3\tau$. It can be see that as the fringes approach synchronisation, the fringes rotate towards the horizontal axis and become more separated.}
\label{fig:MATexp}
\end{figure}

The Astra Gemini laser is fitted with a delay stage which can move in increments of $15$ $\mu$m. This is a double pass delay giving a total change to the length of the beam path of $30$ $\mu$m, which corresponds to a temporal change of $100$ fs i.e., a delay resulting in $\tau=37.5$. A prediction of the behaviour based on equation \ref{eq:specinfsupersimp} is given in figure \ref{fig:MATexp}. 

As can be seen in figure \ref{fig:MATexp}, when the time delay is changed, the fringes both rotate and their separation increases. This occurs as the spatial fringes are not altered by the time delay imposed on the temporal fringes; however, the temporal fringes decouple from each other leading to a compression of those fringes. As each wavelength shows the same number of spatial fringes, the fringes both compress and rotate. At synchronisation, the temporal fringes are no longer visible but spatial fringes are. 

In any arrangement where one is able to steer a two pulses with a small separation to a diffraction grating, and image the 1st order reflection from it, a set of fringes which rotate as the delay between the two pulses is altered should be seen at the CCD. This will allow one to alter the delay between the beams until a clear indication of temporal synchronisation is given. This takes place in the form of horizontal fringes (see figure \ref{fig:MATexp}).

\section{Experimental Setup}\label{sec:Expsetinf}

\begin{figure}[h!]
\centering\includegraphics[width=1.0\textwidth]{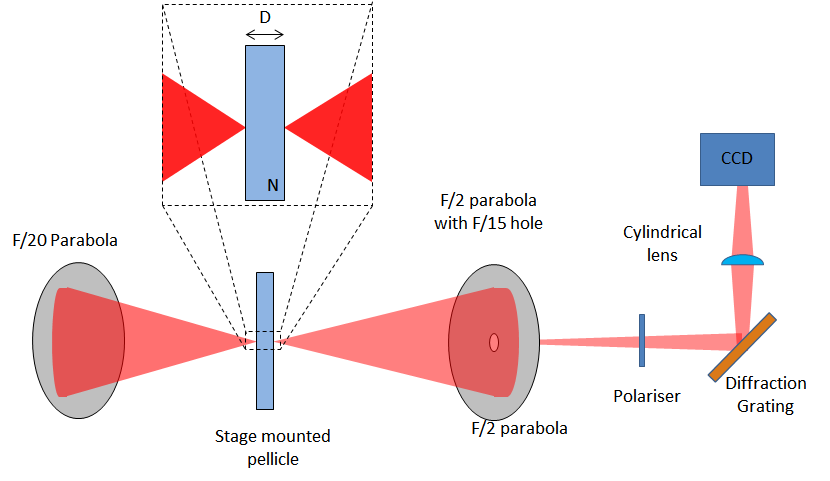}
  \caption{Two counter-propagating laser pulses are incident on the pellicle of thickness $D$ and refractive index $N$. A polariser is used to alter the relative intensities of the beams and allow the beams to interfere with one another. The diffraction grating spreads the beams in terms of frequency in one axis only, while the lens compresses this spread giving greater clarity of signal on the CCD chip. Due to the subtle path differences between the pulses imposed by the pellicle, a unique interference pattern will be formed for each time delay.}
  \label{fig:Experimentalsetup}
\end{figure}

Figure \ref{fig:Experimentalsetup} shows the arrangement used to investigate the technique. After the two counter-propagating laser pulses are spatially overlapped, a pellicle is driven in between the beams in the plane of overlap (the position where timing is critical). A polariser is used to alter the relative intensities of the beams and alter their polarisation. The diffraction grating spreads the beams in terms of frequency in one axis only, while the lens compresses this spread to give greater clarity of signal on the CCD chip. Due to the subtle path differences between the pulses imposed by the pellicle, a unique interference pattern will be formed for each time delay. 

\section{Results}\label{sec:resultsinf}

\begin{figure}[h!]
    \includegraphics[width=1.0\textwidth]{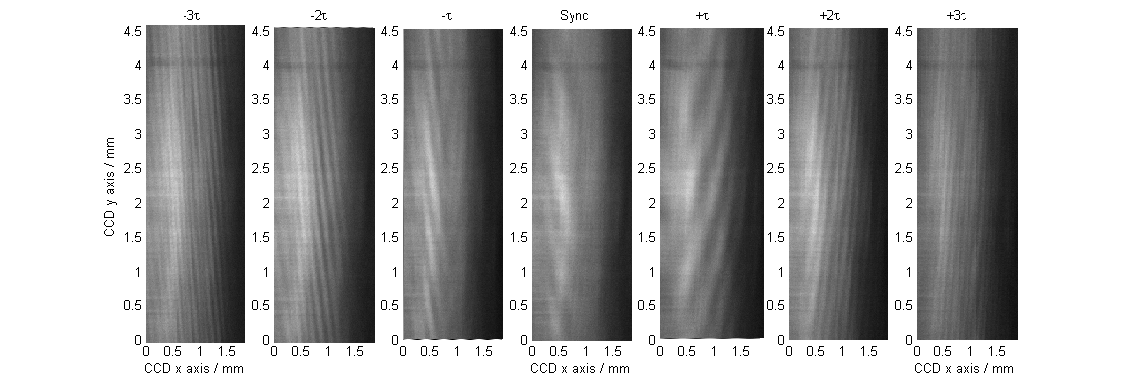}
  \caption{Results of the synchronisation process as seen on the CCD camera. The $7$ images begin from left at a time delay of $-300$ fs progressing to $+300$ fs, in time steps of $100$ fs. It can be seen that the fringes begin almost vertically and rotate towards the horizontal plane. The separation between the fringes also changes. The central image shows the CCD closest to synchronisation. As the images progress, the behaviour of the fringes is mirrored to the corresponding time steps.}
  \label{fig:Infdat}
\end{figure}

Looking at Figure \ref{fig:Infdat}, one can see the changes that occur on the CCD camera close to synchronisation. The $7$ images begin from left at a time delay of $-300$ fs progressing to $+300$ fs, in time steps of $100$ fs. The central image shows the pattern obtained closest to synchronisation. 

Progressing from left to the central image in figure \ref{fig:Infdat}, one can see the fringes begin to rotate towards the horizontal axis. This is within the expectations discussed previously in Section \ref{sec:Theoryspecinf}. Continuing through the images from the central image to right, the behaviour of the fringes is mirrored with the corresponding negative time step to the left, as the delay moves from synchronisation. 

\begin{figure}[h!]
    \includegraphics[width=1.0\textwidth]{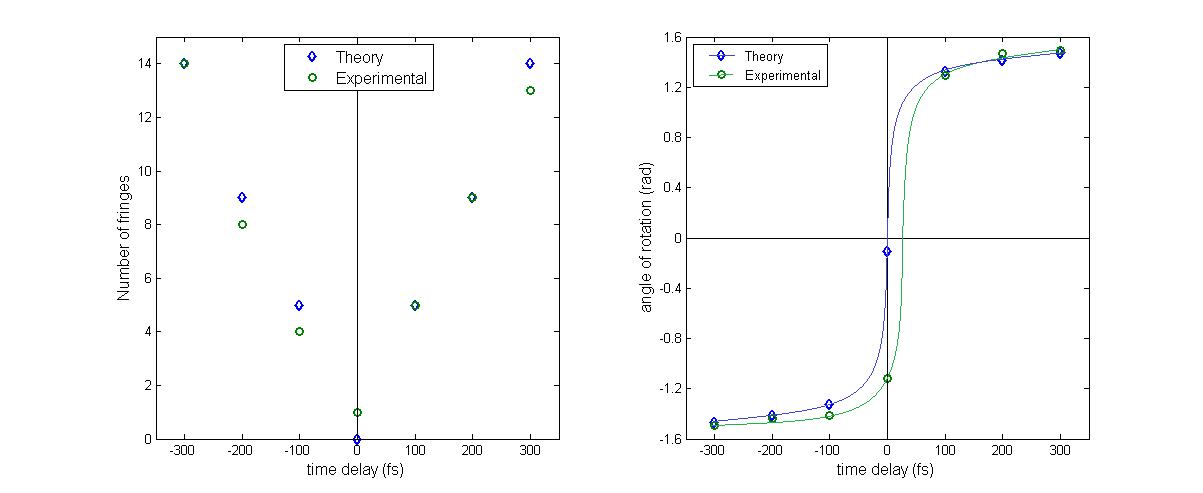}
  \caption{A comparison between the theoretical expectations of the number of temporal fringes predicted (blue) in the model in figure \ref{fig:MATexp} and those obtained experimentally (green) from a lineout at the corresponding on the CCD image. The angles are taken from images on the CCD at time delays of $-300$, $-200$, $-100$ $0$, $+100$, $+200$ and $+300$ fs, respectively.(b) Shows the rotation of fringes during the synchronisation process as seen on the CCD camera at the same delay positions. It can be seen that the rotation of the fringes occurs slowly initially but undergoes a very rapid change. This behaviour is mirrored by the separation of the fringes.  The angles predicted from the model in figure \ref{fig:MATexp} (blue) are compared with those obtained experimentally (green). Lines acting as guides for the eye are provided.Closest to ideal synchronisation, the fringes predicted in figure \ref{fig:MATexp} have an angle of around $0.5$ rad, while the experimental fringes are observed with an angle of $-1.1$ rad.}
  \label{fig:fringes}
\end{figure}

Clearly, synchronisation can be achieved within $\pm50$ fs. Using figures \ref{fig:MATexp} and \ref{fig:Infdat}, a comparison between the number of spatial fringes expected theoretically and those obtained experimentally along the y axis was made. If one counts the number of fringes along the vertical axis at any given position on the horizontal axis in figure \ref{fig:Infdat}, the number obtained is always 4. This is true also for the theoretical model shown in figure\ref{fig:MATexp}. Figure \ref{fig:fringes}(a), shows a similar comparison made between the number of temporal fringes predicted theoretically in figure \ref{fig:MATexp} and obtained experimentally from a fixed point along the x axis. Again there is strong agreement however, some variation is seen. 

Another subtle difference between the images in figure \ref{fig:Infdat} and the predictions made in figure \ref{fig:MATexp} is the rotation of the fringes at synchronisation. Plotting the angle of the fringes with time gives an estimate how far the laser pulses were from ideal synchronisation.

\begin{figure}[h!]
\includegraphics[width=1.0\textwidth]{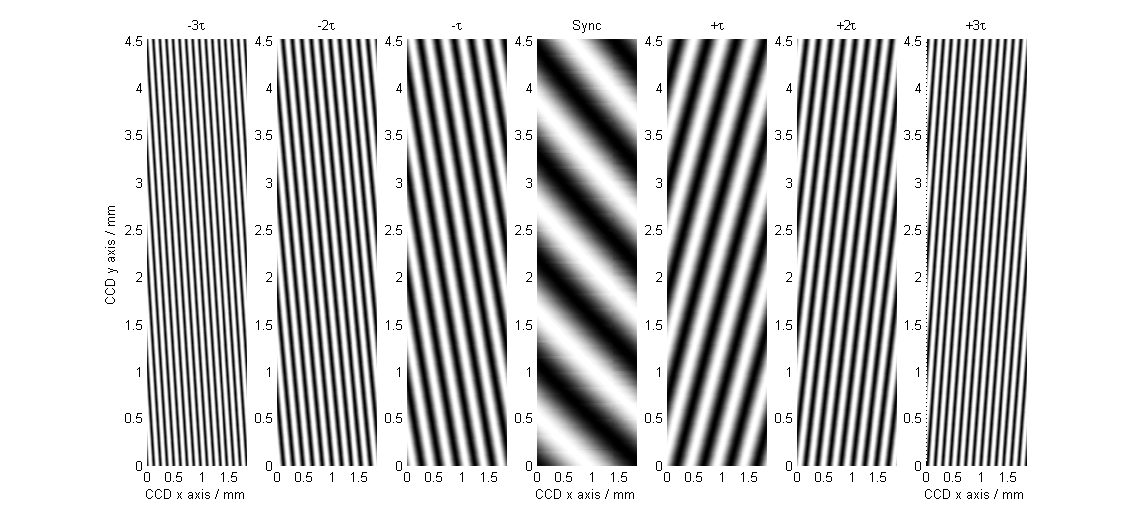} 
  \caption{The relative intensity plotted as a function of the x,y positions with values plotted in term of the phase difference associated with each position on the x-y axis, T is the time normalised motion of the delay stage. Moving from left to right the delays are $-3\tau$, $-2\tau$, $-1\tau$, $0$, $+1\tau$, $+2\tau$ and $+3\tau$, respectively. In this case a systematic error of $-20$ fs is induced in each time change.}
\label{fig:MATexp2}
\end{figure}

The angle of the fringes expected theoretically and those obtained experimentally is given in figure \ref{fig:fringes}(b). From this it can be seen that the rotation of the fringes occurs slowly initially but undergoes a very rapid change. Closest to perfect synchronisation, the fringes obtained experimentally have an angle of around $-1.1$ rad, greater than the $-0.5$ rad predicted in figure \ref{fig:MATexp}. Using the guide for the eyes provided in figure \ref{fig:fringes}(b), this corresponds to a loss in synchronisation of around $20$ fs. This most likely comes from a systematic error in the motion of the delay stage. Imposing the systematic error onto the theoretical predictions made in figure \ref{fig:MATexp}, the agreement with experimental data is enhanced further as can be seen in figure \ref{fig:MATexp2}. 

\section{Conclusions}\label{sec:conclusioninf}
A method for synchronising ultra-short, ultra-intense laser beams in a real-time way, with a precision of $\pm50$ has been successfully demonstrated. It has been shown that the degree of synchronisation achievable, is limited by the step size available on the delay stage and hence, one can easily improve upon the experimental evidence provided here in order to achieve fs scale synchronisation. Due to the compact and versatile nature of the technique, it should prove highly beneficial in reducing the technical difficulties that may arise specifically in the next class of multi-petawatt laser systems such as the 20PW Vulcan at the Rutherford Appleton Laboratory and the Extreme Light Infrastructure, in which experiments with multiple ultra-high intensity lasers will be routinely performed.

\section*{Acknowledgements}
D.J. Corvan and G. Sarri wish to acknowledge the financial support from EPSRC (grant number EP/L013975/1) in addition the authors are grateful to G.Nersisyan and to the technical staff at the Rutherford Appleton Laboratory for the technical support provided.

\bibliographystyle{unsrt}

\begin{thebibliography}{10}

\bibitem{WDM}
D.~Riley, N.~C. Woolsey, D.~McSherry, I.~Weaver, A.~Djaoui, and E.~Nardi.
\newblock X-ray diffraction from a dense plasma.
\newblock {\em {\em Phys. Rev. Lett. \textbf{84} 1704 (2000)}}.

\bibitem{Piazza}
A.~Di~Piazza, C.~M\:{u}ller, K.~Z. Hatsagortsyan, and C.~H. Keitel.
\newblock Extremely high-intensity laser interactions with fundamental quantum
  systems.
\newblock {\em {\em Rev. Mod. Phys. \textbf{84} 1177 (2012)}}.

\bibitem{Labastro}
Bruce~A. Remington, R.~Paul Drake, and Dmitri~D. Ryutov.
\newblock Experimental astrophysics with high power lasers and $z$ pinches.
\newblock {\em {\em Rev. Mod. Phys. \textbf{78} 755 (2006)}}.

\bibitem{2beamacc}
C.~Rechatin L. Ammoura A. Ben-Isma\:{i}l X. Davoine G. Gallot J P. Goddet E.
  Lefebvre V. Malka \& J.~Faure O.~Lundh, J.~Lim.
\newblock Few femtosecond, few kiloampere electron bunch produced by a
  laser–plasma accelerator.
\newblock {\em {\em Nat. Phys. \textbf{7} 219 (2011)}}.

\bibitem{twobeamLWFA}
D.~Umstadter, J.~K. Kim, and E.~Dodd.
\newblock Laser injection of ultrashort electron pulses into wakefield plasma
  waves.
\newblock {\em {\em Phys. Rev. Lett. \textbf{76} 2073 (1996)}}.

\bibitem{bellpairs}
A.~R. Bell and J.~G. Kirk.
\newblock Possibility of prolific pair production with high-power lasers.

\bibitem{ChrisNLTS}
Chris Harvey, Thomas Heinzl, and Anton Ilderton.
\newblock Signatures of high-intensity compton scattering.
\newblock {\em Phys. Rev. A \textbf{79} 063407 (2009)}.

\bibitem{MackNLCS}
F.~Mackenroth and A.~Di Piazza.
\newblock Nonlinear compton scattering in ultrashort laser pulses.
\newblock {\em {\em Phys. Rev. A \textbf{83} 032106 (2011)}}.

\bibitem{specinforig}
M.~Vampouille C.~Froehly J.~Piasecki, B.~Colombeau and J.~A. Arnaud.
\newblock Nouvelle m\'{e}thode de mesure de la r\'{e}ponse impulsionnelle des
  fibres optiques.
\newblock {\em {\em Appl. Opt. \textbf{19} 3749 (1980)}}.

\bibitem{specinfpaper}
D.~Yelin D.~Meshulach and Y.~Silberberg.
\newblock Real-time spatial–spectral interference measurements of ultrashort
  optical pulses.
\newblock {\em {\em Opt. Soc. Am. B \textbf{14} 2095 (1997)}}.

\bibitem{VulcanPW}
C.N. Danson, P.A. Brummitt, R.J. Clarke, J.L. Collier, B.~Fell, A.J.
  Frackiewicz, S.~Hancock, S.~Hawkes, C.~Hernandez-Gomez, P.~Holligan, M.H.R.
  Hutchinson, A.~Kidd, W.J. Lester, I.O. Musgrave, D.~Neely, D.R. Neville, P.A.
  Norreys, D.A. Pepler, C.J. Reason, W.~Shaikh, T.B. Winstone, R.W.W. Wyatt,
  and B.E. Wyborn.
\newblock Vulcan petawatt—an ultra-high-intensity interaction facility.
\newblock {\em {\em Nucl. Fusion \textbf{44} 239 (2004)}}.

\bibitem{ELI}
Extreme~Light Infrastructure.
\newblock "\url{http://www.eli-beams.eu/}".

\bibitem{Geminispec}
C.J.~Hooker \textit{et al}.
\newblock {\em {\em J. Phys. IV France \textbf{133}, 673 (2006)}}.

\end{thebibliography}

\end{document}